\documentclass[journal]{IEEEtran}

\usepackage{cite}
\usepackage[top=0.75in,bottom=1in,left=0.625in,right=0.625in]{geometry}
\usepackage{amsmath,amssymb,amsfonts}
\usepackage{graphicx}
\usepackage{textcomp}
\usepackage{xcolor}
\usepackage{booktabs}
\usepackage{multirow}
\usepackage{float}
\usepackage{microtype}
\usepackage{hyperref}

\hypersetup{colorlinks=true,linkcolor=blue,citecolor=blue,urlcolor=blue}

\begin{document}

\title{Adversarial Reinforcement Learning for Adaptive\\
Eavesdropping in BB84 Quantum Key Distribution}

\author{Ramon Jose C. Bagunu\\
  \small Department of Physical Sciences and Mathematics, University of the Philippines Manila, Ermita, Manila, Philippines\\
  \small \textit{rcbagunu@up.edu.ph}}

\date{}
\maketitle

\begin{abstract}
BB84 quantum key distribution derives its security from a
physical guarantee that any eavesdropper disturbs the channel
in a statistically detectable way. Prior work evaluates this
by assuming Eve attacks at a fixed, analytically optimized
rate. We examine what happens when Eve is modeled instead as
a learning agent. Classical reinforcement learning is used,
specifically tabular Q-Learning, SARSA, and Double Q-Learning,
to adaptive BB84 eavesdropping. This formulates the attacker's
decision as a Markov Decision Process where the agent
observes Quantum Bit Error Rate (QBER) feedback and decides,
qubit by qubit, whether to intercept or pass. Experiments
span three channel noise levels ($\mu_{ch}\in\{1\%,3\%,5\%\}$)
and are validated across five independent random seeds
(45 training runs per condition, 10,000 episodes each).
Against the best non-adaptive analytical baseline, Q-Learning
reduces detection from 99.4\% to $0.28\%\pm0.27\%$ at
$\mu_{ch}=1\%$ while extracting approximately 10.5 correct
bits per episode. This is a 355-fold reduction that is statistically
significant ($p=0.020$, Mann-Whitney $U$ test). We also
report the spontaneous emergence of an \emph{end-game burst},
where agents independently learn to surge their attack rate
at the final block. This exploit vanishes under randomized
checkpoint intervals while stealth performance remains
statistically indistinguishable. These results motivate the
inclusion of adaptive adversary baselines in quantum
cryptographic security evaluations.
\end{abstract}

\medskip

\section{Introduction}

Quantum Key Distribution (QKD) addresses one of the central
problems in secure communication, where we ask how can two parties share a
secret key whose security is guaranteed not by computational
hardness, but by physical laws? The BB84
protocol~\cite{bennett1984quantum}, introduced by Bennett and
Brassard in 1984, is the oldest and most widely deployed
instantiation of this idea. It works by encoding key bits as
photon polarizations in one of two randomly chosen
bases, which are the rectilinear ($+$) and diagonal ($\times$) basis. This relies on the fact that a quantum
state cannot be measured without disturbing
it~\cite{wootters1982single}. Consider an eavesdropper, Eve,
who intercepts a qubit and must guess its basis before
measuring. When she guesses wrong, which happens 50\% of the
time, she irreversibly alters the state before retransmitting
it. This introduces errors at the receiver. If Eve intercepts
every qubit, roughly 25\% of Alice and Bob's bits will
disagree. Legitimate parties detect this by computing the
Quantum Bit Error Rate (QBER) over a sample of their shared
bits, and abort the session if QBER exceeds
$\approx$11\%~\cite{shor2000simple,mayers2001unconditional,
lo1999unconditional}.

The practical complication is that real quantum channels are
noisy independent of Eve. Background channel noise $\mu_{ch}$
flips bits at a rate that is physically indistinguishable from
eavesdropping-induced errors. This provides Eve with operating
room. Any QBER increase below $11\% - \mu_{ch}$ cannot be
reliably attributed to an attack. Statistical fluctuations in
finite block sizes widen this margin
further~\cite{gottesman2004security,scarani2009security}.
Prior work has characterized this margin almost entirely from
the defender's viewpoint, which asks given a fixed attack strategy, how
reliably can it be
detected~\cite{lee2022quantum,gisin2002quantum,xu2020secure}? Complementary to this, we ask, can an adaptive attacker learn to
exploit this margin more effectively than any fixed strategy
allows? This question actually has received almost no attention.

In this paper we study exactly that question. To do so, we
model Eve as a classical tabular RL agent that observes the
QBER announcement at each checkpoint and decides, for every
incoming qubit, whether to intercept or pass. There is no
quantum computing involved, and the quantum mechanics enters only
through the simulation environment. Over 10,000 training
episodes, the agent updates its policy based on the
consequences of its decisions. These include surviving checkpoints, gaining
information, or being detected. This sequential,
feedback-driven framing is precisely the setting that
reinforcement learning is designed to address.

It is also worth noting that the findings here are more
pronounced than one might initially expect from such a simple
setup. At low channel noise, a trained agent achieves
near-perfect stealth not through any sophisticated mechanism
but through temporal restraint. That is, attack conservatively when
future checkpoints remain, and freely at the terminal block
where detection carries no future cost. This \emph{end-game
burst} is not programmed into the reward function, but it emerges
spontaneously in all three algorithms independently. We also
observe that while randomizing checkpoint intervals eliminates
the burst, it does not restore stealth performance. This points to a deeper structural vulnerability than the burst
itself.

We make four contributions:

\begin{enumerate}
  \item We formulate BB84 eavesdropping as a Markov Decision
    Process (MDP) and train three classical tabular RL agents
    to find adaptive attack strategies, validated across five
    independent random seeds per configuration.
  \item We demonstrate that adaptive agents reduce detection
    rates by up to 355-fold versus the best non-adaptive
    baseline, with differences statistically significant across
    multiple noise conditions (Mann-Whitney $U$ test,
    $p<0.05$).
  \item We report the spontaneous emergence of an
    \emph{end-game burst}, which is a temporal exploit of BB84's
    fixed-checkpoint structure discovered independently by all
    three algorithms. We confirm experimentally that it
    vanishes under randomized checkpoint intervals.
  \item We derive concrete defensive recommendations directly
    from the observed attack behaviors, including why
    checkpoint randomization, while beneficial, is not
    sufficient on its own.
\end{enumerate}

We emphasize that this work is a proof of concept. The goal
is to establish that adaptive eavesdropping strategies are
learnable in principle under the BB84 reward structure, not
to engineer a fully realistic eavesdropping system. The
simplicity of the setup is deliberate because we use the most basic
classical RL agents in the most transparent possible
environment. This ensures that the findings reflect the structure
of the BB84 protocol itself, rather than the sophistication
of the learning algorithm.

\section{Related Work}

This work sits at the intersection of two bodies of
literature. First is the QKD security analysis which characterizes the
theoretical limits of eavesdropping detection, and second is the reinforcement learning for security which provides the
adaptive attacker model we employ. We review both and clarify
how our use of classical RL relates to recent quantum machine
learning approaches to QKD.

\subsection{BB84 Security Analysis}

BB84's unconditional security under ideal conditions was proven
independently by Shor and Preskill~\cite{shor2000simple},
Mayers~\cite{mayers2001unconditional}, and Lo and
Chau~\cite{lo1999unconditional}. Gottesman et
al.~\cite{gottesman2004security} extended these proofs to
cover imperfect devices. Comprehensive reviews of practical
QKD security appear in Scarani et
al.~\cite{scarani2009security}, Xu et al.~\cite{xu2020secure},
and Gisin et al.~\cite{gisin2002quantum}. Brassard et
al.~\cite{brassard2000limitations} characterized
photon-number-splitting attacks on weak coherent sources.
In all of this prior work, it can be observed that Eve's
strategy is modeled as fixed, either full intercept-resend
or a specified probabilistic variant. The question of whether
an adaptive Eve observing QBER feedback can do substantially
better has not been studied.

The work most directly relevant to ours is Lee et
al.~\cite{lee2022quantum}, who derive the analytically optimal
fixed attack rate $r^* = (0.11 - 0.005 - \mu_{ch})/0.25$
and study QBER-based detection under this strategy. Their
model serves as our primary baseline. The key limitation of
the fixed-rate assumption is that it ignores the QBER
announcements that Alice and Bob broadcast at each
checkpoint. This is a signal that an adaptive attacker can use to
modulate attack intensity in response to accumulated risk.
Our work examines what happens when Eve actually uses that
information.

\subsection{Reinforcement Learning in Security and QKD}

Reinforcement learning has been applied to network intrusion
detection~\cite{nguyen2023deep} and routing in QKD-secured
optical networks~\cite{sharma2023deep}. In the quantum domain,
Rei\ss{} and van Loock~\cite{reiss2023deep} optimize quantum
repeater strategies via deep RL, and Decker et
al.~\cite{decker2024qkd} frame QKD itself as a quantum machine
learning task. Of particular note is Behera et
al.~\cite{behera2025qnnqrl}, who integrate quantum neural
networks with quantum RL circuits running on quantum hardware
to improve BB84 key generation accuracy, which effectively
optimized the communicants' side of the protocol.

It is worth contrasting our approach with these quantum ML
works. The agents used here are standard classical tabular
algorithms from the
1990s~\cite{watkins1992q,rummery1994line,hasselt2010double}, methods
with well-understood convergence properties, no quantum
hardware requirement, and inspectable learned policies.
One can read Q-values for every state and trace precisely why
the agent attacks when it does. This level of transparency
is not straightforwardly available in quantum circuit-based
approaches, and it is precisely what allows us to attribute
the end-game burst to protocol structure rather than a
training artifact. Our contribution is also directionally
orthogonal to Behera et al. They optimize the defender's
side of the protocol, while we optimize the attacker's.

Concurrent work has examined adaptive defenders against
side-channel attacks in decoy-state QKD via minimax
learning~\cite{mohamed2026adversarial} and the limits of
quantum certification under GAN-based adversaries~\cite{tasar2025adversarial}. 
Our contribution differs in studying the attacker side of standard BB84 rather than defender-side detection in decoy-state systems or certification-level adversarial bounds.
\section{Methodology}

We now describe the simulation environment, the MDP
formulation, the three agent implementations, and the
experimental protocol. The overarching design choice
throughout is transparency. We have a straightforward simulation,
a small discrete state space, and classical tabular algorithms
whose behavior can be fully traced and inspected.

\subsection{BB84 Simulation}

We implement a qubit-level simulation following Lee et
al.~\cite{lee2022quantum}. Each episode consists of 1,000
qubits organized into 10 blocks of 100 qubits each. Alice
randomly generates a bit and basis for each qubit and encodes
it as a photon polarization. Eve optionally intercepts by
measuring in a randomly chosen basis. If her basis matches
Alice's (probability 0.5), she learns the correct bit with
no disturbance. Otherwise, her measurement collapses the
qubit into a random state, which introduces a 25\% error into
Bob's subsequent measurement. The channel independently
flips any qubit with probability $\mu_{ch}$. At the end of
each block, Alice and Bob sift their bits (retaining only
those where their bases matched) and compute QBER over all
sifted bits. If QBER $\geq 11\%$, the session is aborted.

Four sanity checks that confirm physical correctness are the following: (i)~0\% QBER with no noise and no attacks; (ii)~$\approx$25\% QBER
under full attack with no noise; (iii)~QBER
$\approx \mu_{ch}$ with noise only; (iv)~QBER matching the
Lee et al.\ analytic formula under partial attack rates. All
checks pass within $\pm$0.02\%.

\subsection{MDP Formulation}

Eve's decision problem is formulated as an MDP
$\langle\mathcal{S}, \mathcal{A}, \mathcal{R}, \mathcal{T},
\gamma\rangle$. After each checkpoint, Eve observes (1)~the
announced QBER, (2)~her attack count in the current block,
and (3)~the block index. These are discretized into 16 QBER
bins over $[0\%, 15\%]$ in steps of 1\%, 21 attack bins over
$[0, 100]$ in steps of 5, and 10 block index bins (0--9),
which yields a total of $16 \times 21 \times 10 = 3{,}360$ states.
The action set is $a_t \in \{0, 1\}$, which means pass or attack, decided once per qubit.

Eve receives per-qubit feedback and a checkpoint-level signal
at each block boundary, given by
\begin{equation}
r_t = \begin{cases}
+1  & \text{correct basis (information gained)} \\
-1  & \text{wrong basis (noise introduced, no gain)} \\
+2  & \text{checkpoint survived} \\
-50 & \text{detected (QBER} \geq 11\%) \\
0   & \text{pass (no attack).}
\end{cases}
\label{eq:reward}
\end{equation}

It can be seen that the $-1$ wrong-basis penalty sets the
expected immediate reward of attacking to zero, which forces
the agent to learn when to attack rather than simply
attacking at every opportunity. The $+2$ survival bonus
creates an incentive for multi-block restraint. The $-50$
detection penalty dominates all other signals, which ensures the
agent prioritizes evasion.

\subsection{Agent Implementations}

All three agents are standard classical tabular methods
sharing identical infrastructure where we have $\varepsilon$-greedy
exploration with $\varepsilon$ decaying linearly from 1.0 to
0.01 over 10,000 episodes. Suppose random exploration is left unbiased
at a 50\% attack rate. In this case, the resulting QBER
pushes above threshold in nearly every early episode, which causes
universal early detection before any meaningful learning can
occur. We address this by biasing random exploration toward
20\% attack rate, which provides sufficient survival time
for the agent to observe the reward structure. The three
algorithms differ only in their Q-value update rules.

\textbf{Q-Learning}~\cite{watkins1992q} updates off-policy
using the maximum future value, given by
\begin{equation}
Q(s,a) \leftarrow Q(s,a) + \alpha\bigl[r + \gamma\max_{a'}
Q(s',a') - Q(s,a)\bigr].
\end{equation}
The $\max$ operator systematically overestimates action
values, which biases the agent toward more aggressive attacks.

\textbf{SARSA}~\cite{rummery1994line} updates on-policy
using the value of the action actually taken, where the update is
\begin{equation}
Q(s,a) \leftarrow Q(s,a) + \alpha\bigl[r + \gamma Q(s',a')
- Q(s,a)\bigr].
\end{equation}
Because exploration-phase detections feed directly into
value estimates via on-policy updates, SARSA learns a more
conservative policy than Q-Learning.

\textbf{Double Q-Learning}~\cite{hasselt2010double} maintains
two Q-tables, one to select actions and the other to
evaluate them. This corrects Q-Learning's overestimation bias where the update is
\begin{equation}
\begin{split}
Q_A(s,a) \leftarrow Q_A(s,a) + \alpha\bigl[r
  &+ \gamma Q_B(s', \arg\max_{a'} Q_A(s',a')) \\
  &- Q_A(s,a)\bigr].
\end{split}
\end{equation}
Action selection uses the average of both tables. By most
accurately estimating action values, Double Q-Learning
produces the most conservative attack policy of the three.

\subsection{Hyperparameter Selection}

We swept $\alpha \in \{0.01, 0.1, 0.5\}$ and
$\gamma \in \{0.5, 0.75, 0.95, 0.99\}$ for 2,000 episodes
per agent-noise combination (108 total runs) and selected
the combination maximizing mean reward over the final 500
episodes. Table~\ref{tab:hyperparams} summarizes the results.

It can be observed that Q-Learning and Double Q-Learning
consistently prefer $\alpha{=}0.5,\,\gamma{=}0.5$ across all
noise levels. Fast updates and a short planning horizon suit
a setting dominated by the large immediate detection penalty.
SARSA consistently prefers $\gamma{=}0.99$ with a
noise-adapted $\alpha$, which reflects its need to propagate
detection costs backward across many steps via on-policy
updates. The sensitivity is substantial. At $\mu_{ch}=1\%$,
the best and worst Q-Learning combinations differ by 51
reward points over the final 500 episodes.

\begin{table}[t]
\centering
\caption{Selected hyperparameters and mean episode reward
(last 500 episodes of sensitivity sweep).}
\label{tab:hyperparams}
\small
\begin{tabular}{llccr}
\toprule
Agent & $\mu_{ch}$ & $\alpha$ & $\gamma$ & Reward \\
\midrule
\multirow{3}{*}{Q-Learning}    & 1\% & 0.50 & 0.50 & $+$17.15 \\
                                & 3\% & 0.50 & 0.50 & $-$3.62  \\
                                & 5\% & 0.50 & 0.50 & $-$34.96 \\
\midrule
\multirow{3}{*}{SARSA}         & 1\% & 0.01 & 0.99 & $+$11.97 \\
                                & 3\% & 0.10 & 0.99 & $-$5.82  \\
                                & 5\% & 0.50 & 0.99 & $-$36.51 \\
\midrule
\multirow{3}{*}{Double Q-Lrn.} & 1\% & 0.50 & 0.50 & $+$7.62  \\
                                & 3\% & 0.50 & 0.50 & $-$16.86 \\
                                & 5\% & 0.50 & 0.75 & $-$42.52 \\
\bottomrule
\end{tabular}
\end{table}

\subsection{Baselines}

To contextualize the RL results, we implement two non-adaptive
baselines that bound the performance of any non-learning Eve.

\textbf{Always Attack} intercepts every qubit, which yields 100\%
detection at every noise level with sessions aborted at the
first checkpoint. It establishes the ceiling on information
gain, which is approximately 50 correct bits per 1,000-qubit session.

\textbf{Fixed Rate}~\cite{lee2022quantum} attacks each qubit
independently with probability
$r^* = (0.11 - 0.005 - \mu_{ch})/0.25$, which is the analytically
optimal non-adaptive rate. Despite being designed to keep
expected QBER just below threshold, this baseline is detected
99.4\%, 99.6\%, and 98.8\% of the time at $\mu_{ch}=1\%,
3\%,$ and 5\% respectively. It can be seen that statistical
fluctuations in finite 100-qubit blocks push the observed
QBER above 11\% even when the expected value is 10.5\%.
Average episode lengths of 244--272 qubits indicate that
detection typically occurs at the second or third checkpoint.

Taken together, these two baselines define the performance of any non-adaptive strategy. Reckless attack
maximizes information but guarantees detection, and the
analytically optimal fixed rate attempts stealth but fails
nearly 100\% of the time. Neither baseline can exploit QBER
feedback. This is precisely the gap that adaptive RL is
designed to address.

\subsection{Multi-Seed Experimental Protocol}

To produce statistically reliable results, each agent-noise
combination is trained five times with independent random
seeds (42, 123, 456, 789, 1337). All reported metrics are
mean $\pm$ standard deviation over the final 500 episodes
across these five seeds (45 total runs per condition).
Pairwise agent comparisons use the two-sided Mann-Whitney $U$
test at $\alpha = 0.05$, which is chosen for its suitability with
small samples ($N = 5$) and its non-parametric nature. All
figures in the Results section are generated from these
multi-seed data. Figures showing per-block attack profiles
also overlay per-seed means as scatter points so that the
agreement of independent seeds is directly visible.

\section{Results}

We now present the experimental results, organized in five
parts. We first verify that genuine learning occurs across
all agent-noise combinations before comparing against
baselines. We then assess whether the observed differences
between agents are statistically significant. Following this,
we characterize the stealth--information trade-off, and
finally examine the end-game burst and confirm its structural
origin through a controlled randomization experiment.

\subsection{All Agents Learn to Evade Detection}

\begin{figure}[t]
\centering
\includegraphics[width=\columnwidth]{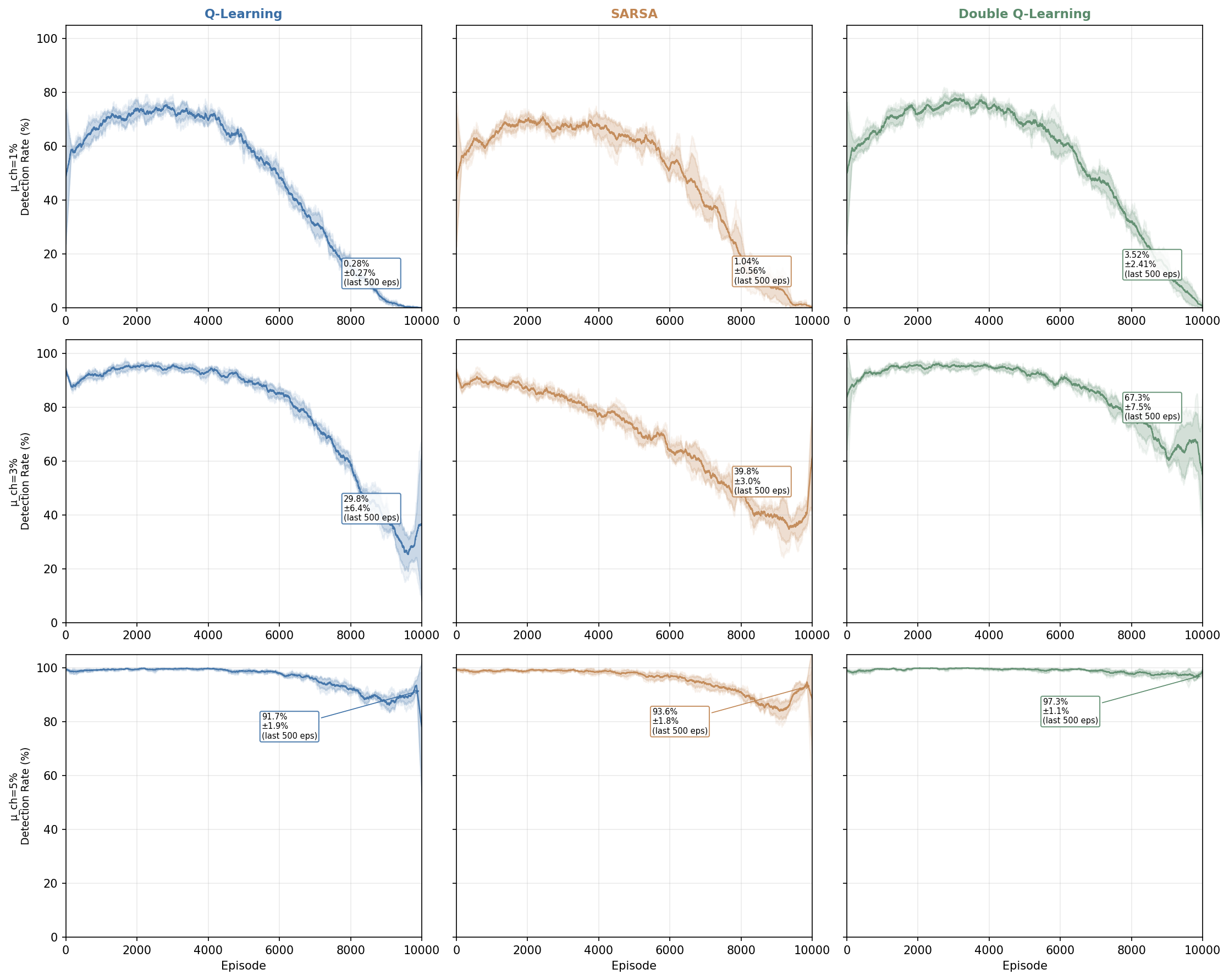}
\caption{Learning curves with shaded $\pm$std bands across
5 seeds. Rows: $\mu_{ch}=1\%, 3\%, 5\%$. Columns:
Q-Learning, SARSA, Double Q-Learning. Annotations report
mean $\pm$ std from the final 500 episodes.}
\label{fig:learning_shaded}
\end{figure}

Fig.~\ref{fig:learning_shaded} shows detection rate
trajectories with shaded $\pm$std bands across seeds. It can
be observed that every agent at every noise level follows
the same broad arc. That is, early episodes are dominated by random
exploration at high detection rates, followed by a clear
and consistent descent as $\varepsilon$ decays and the
learned policy takes over. The narrow variance bands confirm
that this trajectory is reproducible across seeds, and learning
is not a single-seed artifact.

The descent is most dramatic at $\mu_{ch}=1\%$. Q-Learning
drops from approximately 58\% detection at episode 1,000
to $0.28\%\pm0.27\%$ by the final 500 episodes, while mean
episode reward correspondingly improves from $-16.1$ to
$+20.2$. This confirms genuine policy improvement rather than
a reward-shaping artifact. At $\mu_{ch}=5\%$, by contrast,
all agents plateau above 88\% detection with persistently
negative rewards ($-37$ to $-47$). The safe attack budget
at this noise level, roughly 22 qubits per 100, is
insufficient for any profitable strategy under the current
reward structure.

\subsection{RL Agents Outperform Baselines}

\begin{figure}[t]
\centering
\includegraphics[width=\columnwidth]{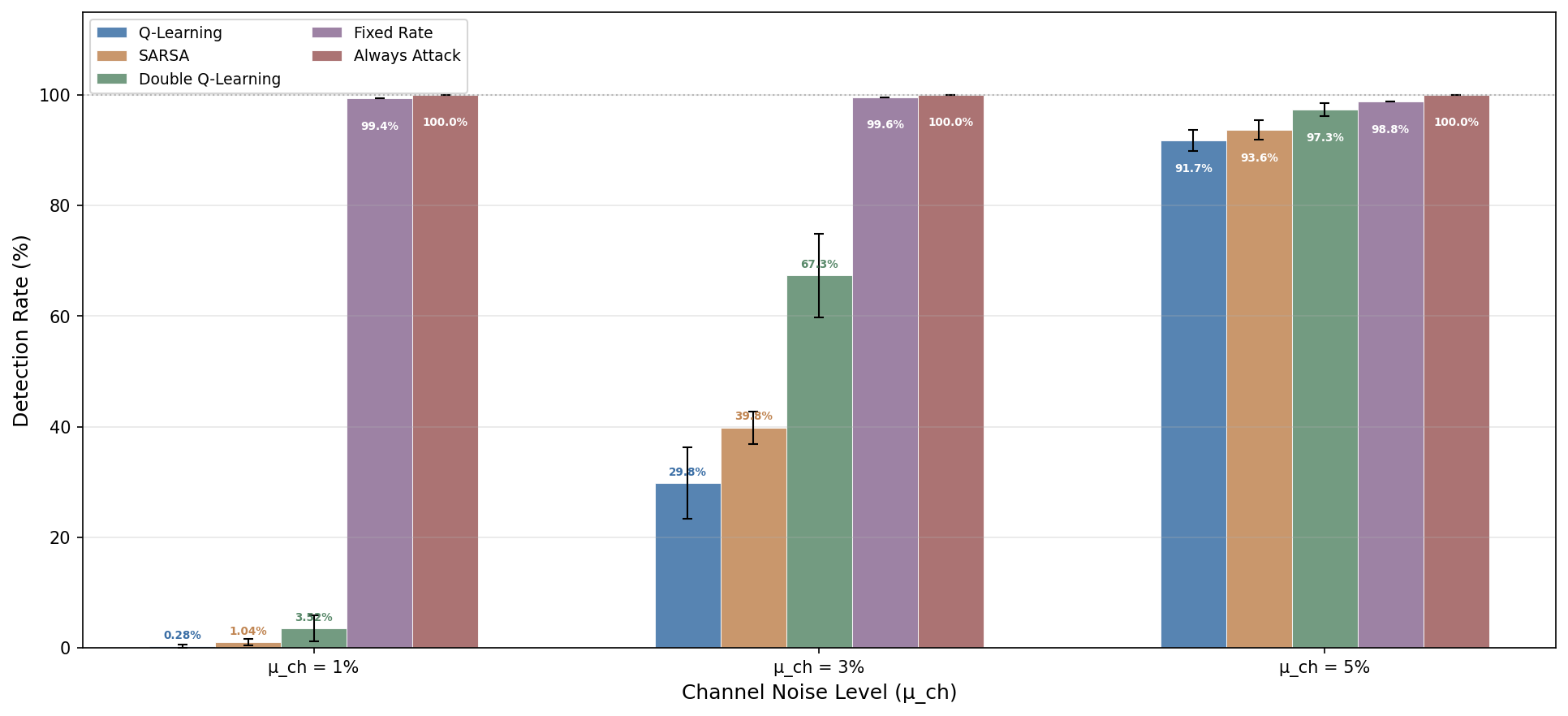}
\caption{Detection rates for all agents and baselines across
three noise levels (5-seed mean $\pm$ std, error bars on the
RL agents, and baselines are deterministic in this metric and
shown without error bars). RL agents (left three bars per
group) achieve near-zero detection at $\mu_{ch}=1\%$, while
both baselines (right two bars) remain near 100\% across all
conditions.}
\label{fig:comparison}
\end{figure}

\begin{figure}[t]
\centering
\includegraphics[width=\columnwidth]{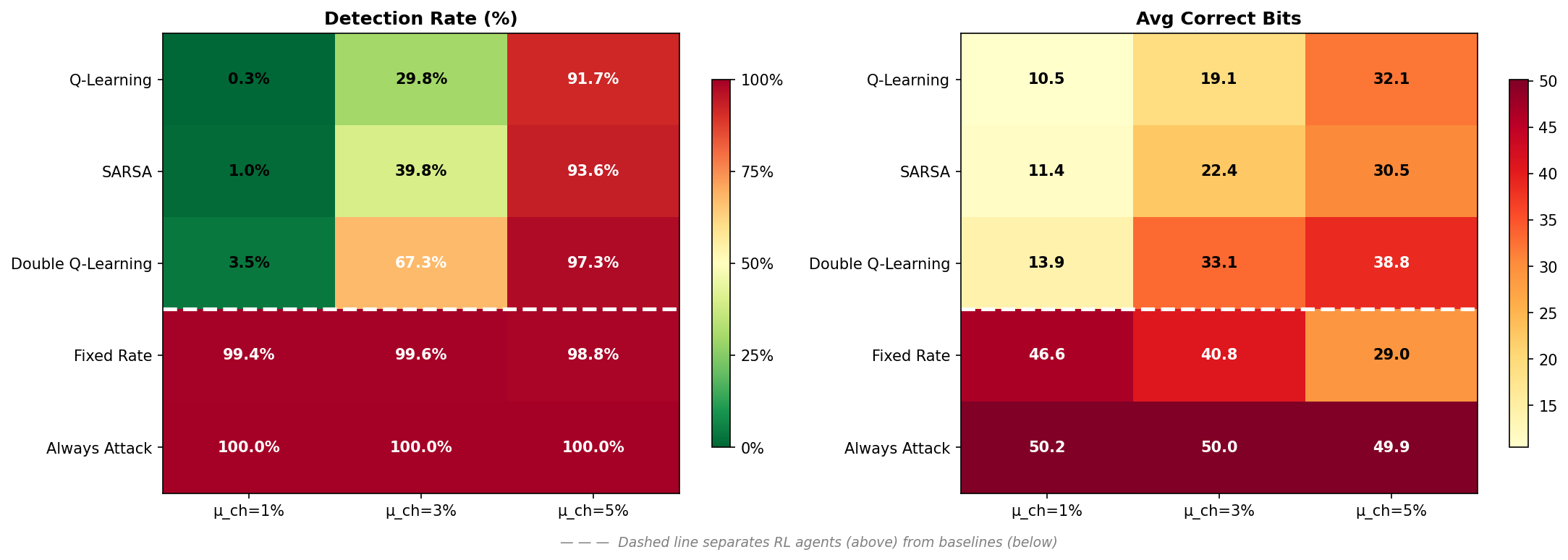}
\caption{Detection rate and average correct bits heatmap for
all agents and noise levels (5-seed mean). RL agents (above
dashed line) consistently occupy the low-detection region
while baselines (below dashed line) cluster near 100\%.}
\label{fig:heatmap}
\end{figure}

Table~\ref{tab:main} reports final detection rates and
correct bits for all agents and baselines.
Fig.~\ref{fig:comparison} and Fig.~\ref{fig:heatmap}
visualize these results. Consider the results at
$\mu_{ch}=1\%$. Q-Learning achieves $0.28\%\pm0.27\%$
detection versus 99.4\% for the Fixed Rate baseline, which is a
355-fold reduction. It is worth recalling that the Fixed
Rate baseline represents the analytical optimum for any
non-adaptive strategy~\cite{lee2022quantum}. An agent that
does nothing beyond observing QBER feedback and updating a
lookup table reduces detection by nearly three orders of
magnitude.

SARSA and Double Q-Learning reduce detection by 96-fold and
28-fold respectively at $\mu_{ch}=1\%$. The ordering
Q-Learning $<$ SARSA $<$ Double Q-Learning by detection rate
is consistent across most conditions and maps directly onto
the theoretical overestimation hierarchy of the three
algorithms. Q-Learning's off-policy $\max$ operator
overestimates future value, thus encouraging more aggressive
attacks. Double Q-Learning corrects this bias most
thoroughly, which produces the most conservative policy. SARSA
sits between them, where its on-policy updates incorporate
exploration-phase detections into value estimates but do not
fully suppress overestimation.

\begin{table*}[t]
\centering
\caption{Detection rate (\%) mean $\pm$ std and avg.\
correct bits across 5 seeds (final 500 episodes).
Bold = lowest detection per noise level.}
\label{tab:main}
\small
\setlength{\tabcolsep}{5pt}
\begin{tabular}{lcccccc}
\toprule
& \multicolumn{3}{c}{Detection Rate (\%)}
& \multicolumn{3}{c}{Correct Bits} \\
\cmidrule(lr){2-4}\cmidrule(lr){5-7}
Agent & $\mu_{ch}=1\%$ & $\mu_{ch}=3\%$ & $\mu_{ch}=5\%$
      & $\mu_{ch}=1\%$ & $\mu_{ch}=3\%$ & $\mu_{ch}=5\%$ \\
\midrule
\textbf{Q-Learning}
  & $\mathbf{0.28\pm0.27}$ & $\mathbf{29.8\pm6.4}$ & $\mathbf{91.7\pm1.9}$
  & $10.5\pm0.3$ & $19.1\pm2.3$ & $32.1\pm0.4$ \\
SARSA
  & $1.04\pm0.56$ & $39.8\pm3.0$ & $93.6\pm1.8$
  & $11.4\pm2.1$ & $22.4\pm1.0$ & $30.5\pm0.6$ \\
Double Q-Learning
  & $3.52\pm2.41$ & $67.3\pm7.5$ & $97.3\pm1.1$
  & $13.9\pm1.4$ & $33.1\pm2.8$ & $38.8\pm0.9$ \\
\midrule
Fixed Rate  & $99.4$ & $99.6$ & $98.8$ & $46.6$ & $40.8$ & $29.0$ \\
Always Attack & $100$ & $100$ & $100$ & $50.2$ & $50.0$ & $49.9$ \\
\bottomrule
\end{tabular}
\end{table*}

\subsection{Statistical Significance}

\begin{figure}[b]
\centering
\includegraphics[width=\columnwidth]{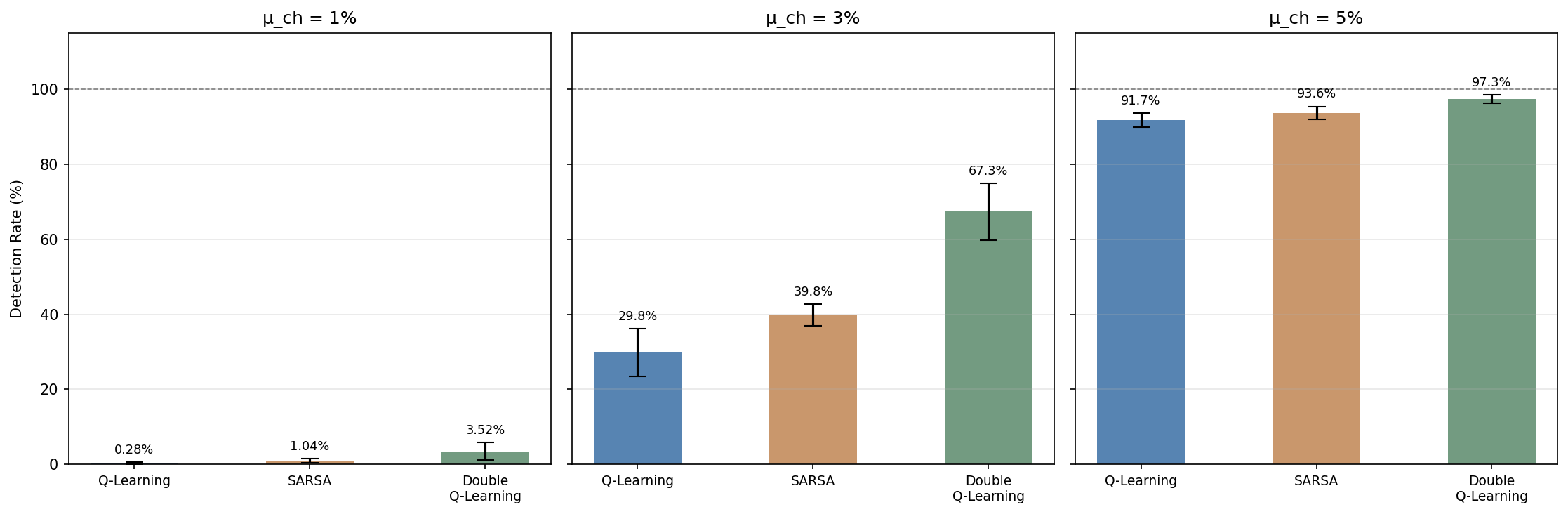}
\caption{Detection rate mean $\pm$ std across 5 seeds. The
tight variance at $\mu_{ch}=1\%$ confirms that Q-Learning's
near-zero detection is robust across seeds rather than a
single-run artifact.}
\label{fig:detection_errorbars}
\end{figure}

\begin{figure}[t]
\centering
\includegraphics[width=\columnwidth]{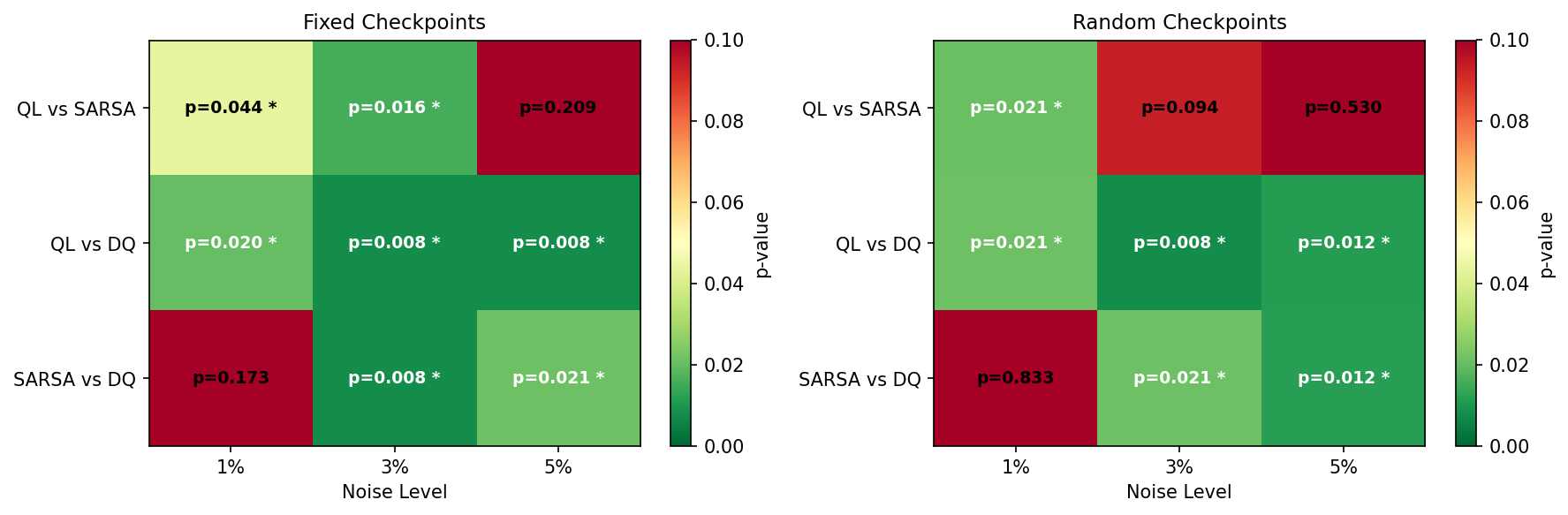}
\caption{Mann-Whitney $U$ test $p$-values for pairwise agent
comparisons (green = significant at $\alpha=0.05$). Left:
fixed checkpoints. Right: random checkpoints. The pattern
of significant and non-significant results is mostly consistent
across both conditions, which suggests that the agent
ranking is not an artifact of protocol structure.}
\label{fig:sig_heatmap}
\end{figure}

Let us now examine the statistical significance of the
observed differences. Fig.~\ref{fig:detection_errorbars}
and Fig.~\ref{fig:sig_heatmap} together provide the full
picture. At $\mu_{ch}=1\%$, Q-Learning significantly
outperforms both SARSA ($p=0.044$) and Double Q-Learning
($p=0.020$). At $\mu_{ch}=3\%$, all pairwise comparisons
reach significance ($p\leq0.016$).

Two comparisons do not reach significance, and it is worth
noting that both are interpretable as meaningful null results.
At $\mu_{ch}=5\%$, Q-Learning and SARSA are statistically
indistinguishable ($p=0.209$). Both achieve $>$88\%
detection in a regime where channel noise dominates strategy,
and a 1.9 percentage-point difference carries no practical
consequence. Similarly, SARSA vs.\ Double Q-Learning at
$\mu_{ch}=1\%$ ($p=0.173$) reflects the fact that both
achieve near-zero detection. Neither is meaningfully superior
in the regime where all agents succeed. These non-significant
results are consistent with theoretical expectations that they
occur precisely where such tests should find no
meaningful difference.

\subsection{The Stealth--Information Trade-off}

\begin{figure}[b]
\centering
\includegraphics[width=\columnwidth]{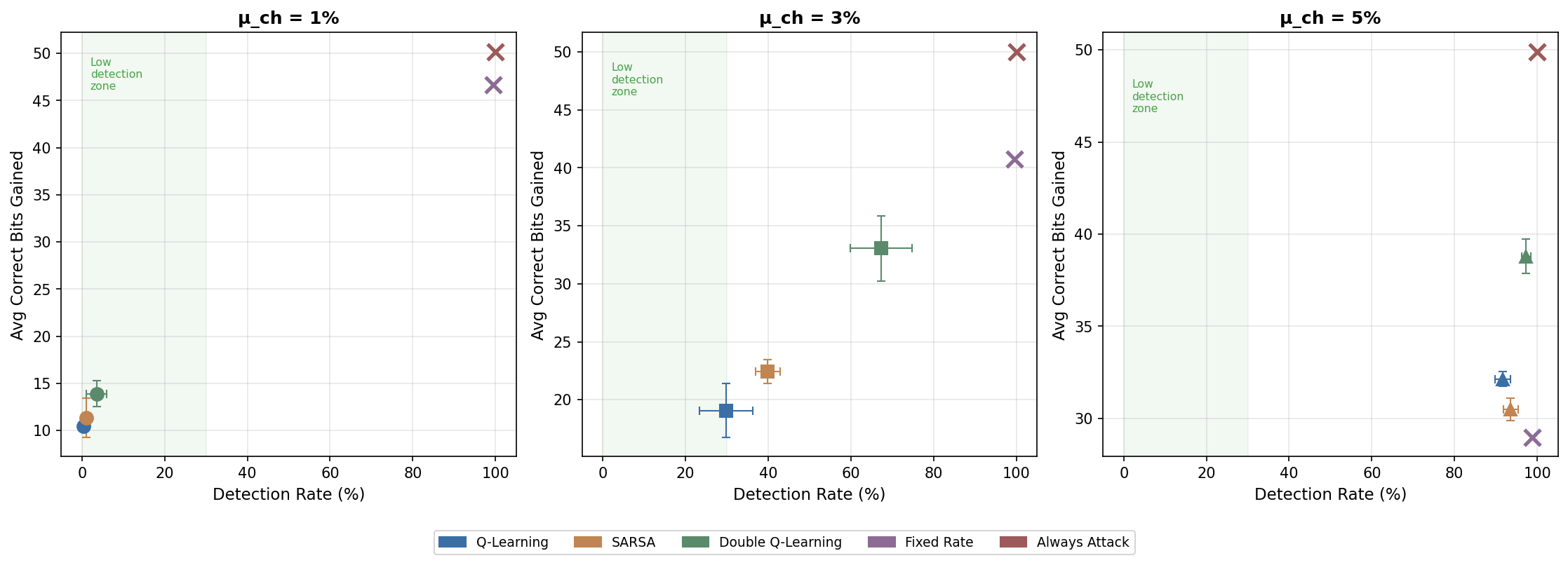}
\caption{Information gain vs.\ stealth trade-off across all
agents and noise levels (5-seed mean, with horizontal and vertical
bars showing $\pm$std across seeds on the RL agents). RL agents
cluster in the low-detection zone while both baselines
(crosses) lie far from it.}
\label{fig:tradeoff}
\end{figure}

An observation that can be drawn from the results is that
lower detection consistently comes at the cost of less
information gathered. Consider Q-Learning at $\mu_{ch}=1\%$. It
gains 10.5 correct bits per episode while remaining
near-undetected, compared to the Fixed Rate baseline's 46.6
bits at 99.4\% detection. Maintaining stealth requires
infrequent attacks, which limits information gain.
Fig.~\ref{fig:tradeoff} makes this trade-off explicit across
all noise levels, where RL agents occupy the bottom-left of the
detection vs.\ bits plane and baselines occupy the
top-right.

The trade-off sharpens with noise. At $\mu_{ch}=3\%$, the
safe attack budget shrinks and agents collect fewer bits,
though they still substantially outperform baselines on
stealth. At $\mu_{ch}=5\%$, the budget is largely exhausted
and all agents fail to maintain stealth and converge toward
detection rates comparable to the Fixed Rate baseline.
The 10.5 correct bits per episode at $\mu_{ch}=1\%$ is
modest in isolation, but it represents sustained, undetectable
leakage across sessions that the 11\% QBER threshold would
never flag.

\subsection{The End-Game Burst}

\begin{figure}[t]
\centering
\includegraphics[width=\columnwidth]{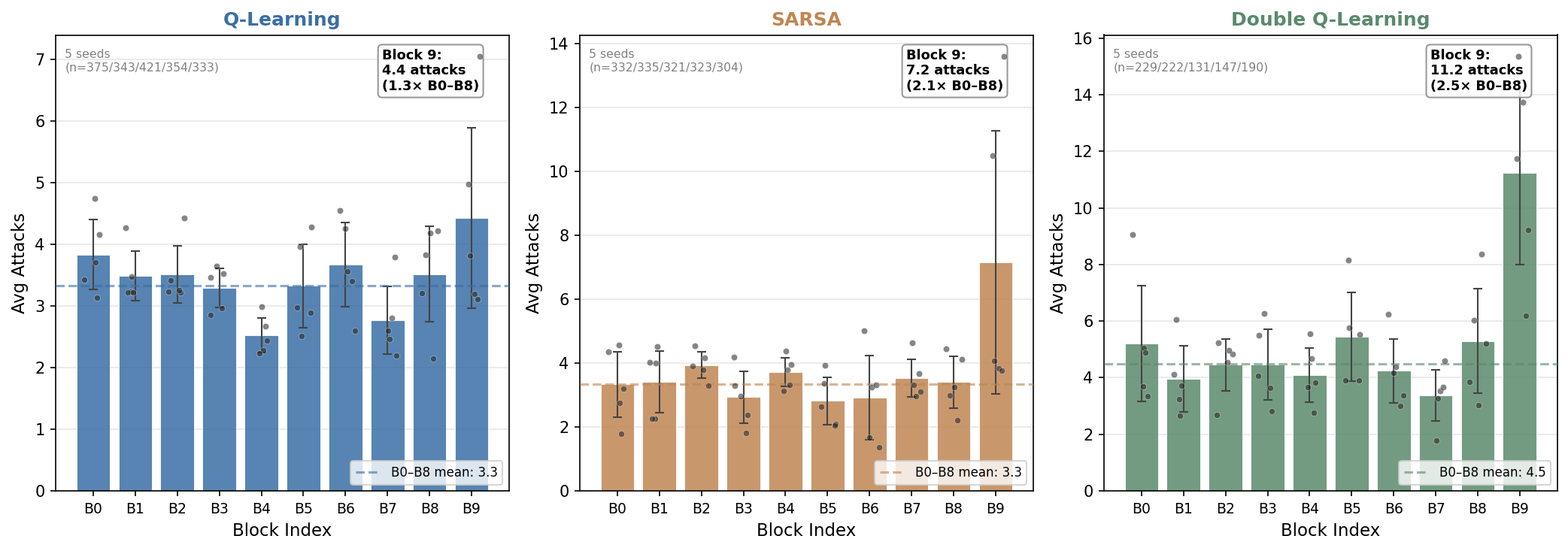}
\caption{Mean attacks per block at $\mu_{ch}=3\%$ (last 500
undetected episodes, full episodes only). Bars show the
5-seed mean with $\pm$std error bars, and the dark scatter points
overlay each seed's mean. The dashed line marks the B0--B8 mean for
each agent. Every agent's block-9 mean exceeds its B0--B8
mean ($1.3\times$ for Q-Learning, $2.1\times$ for SARSA,
$2.5\times$ for Double Q-Learning), with the surge most
consistent across seeds for SARSA and Double Q-Learning.}
\label{fig:burst}
\end{figure}

The most structurally revealing finding is a temporal attack
pattern that emerges in all three agents without any explicit
programming. Fig.~\ref{fig:burst} shows mean attacks per
block over the final 500 undetected episodes at
$\mu_{ch}=3\%$, averaged across all five seeds, with each
seed's mean overlaid as a scatter point. Across blocks 0--8,
attack rates are near-uniform and conservative. At block 9
they rise above the B0--B8 mean for every agent and every
seed, with Q-Learning reaching $4.4$ vs.\ $3.3$ attacks
($1.3\times$), SARSA $7.2$ vs.\ $3.3$ ($2.1\times$), and
Double Q-Learning $11.2$ vs.\ $4.5$ ($2.5\times$). The effect
is most pronounced in Double Q-Learning, which carries the
largest sustained attack rate across blocks 0--8 and also
deposits the largest absolute count at the terminal block.
The per-seed scatter shows the surge is present in every
seed, though its magnitude varies, which is especially
apparent in the Double Q-Learning panel where individual
seeds range from roughly $8$ to $14$ block-9 attacks. The
ordering of burst magnitude across the three agents does not
follow the detection-rate ordering of Section~4.2 and we
discuss this disconnect below.

To understand why this behavior emerges, let's consider the episode
structure. Block 9 is the final block, and detection there
terminates an episode that would end regardless. The
expected future cost of detection at block 9 is therefore
zero. A rational agent will maximize information at this
terminal boundary without constraint. Earlier blocks, by
contrast, carry real cost. Early detection forfeits all
remaining checkpoint-survival rewards and future attack
opportunities. The agent learns this asymmetry through
reward signals alone. It is notable that all three
algorithms independently discover this pattern despite their
substantially different update rules. This shows that
it is a property of the reward landscape rather than an
artifact of any particular algorithm. The fact that
Double Q-Learning, the most conservative agent during
blocks 0--8, also produces the largest terminal surge is
consistent with this account, because a more accurate value
estimate during the constrained portion of the episode leaves
a larger residual budget to be cashed in at the terminal
block where the constraint vanishes.

To test whether this behavior is intrinsic to the agents or
a consequence of the fixed-horizon protocol design, we ran a
controlled experiment where block count is drawn uniformly
from $[5, 15]$ per episode. Under this condition, Eve knows
how many blocks she has survived but not how many remain, which removes the terminal-block signal the burst exploits.

\begin{figure}[t]
\centering
\includegraphics[width=\columnwidth]{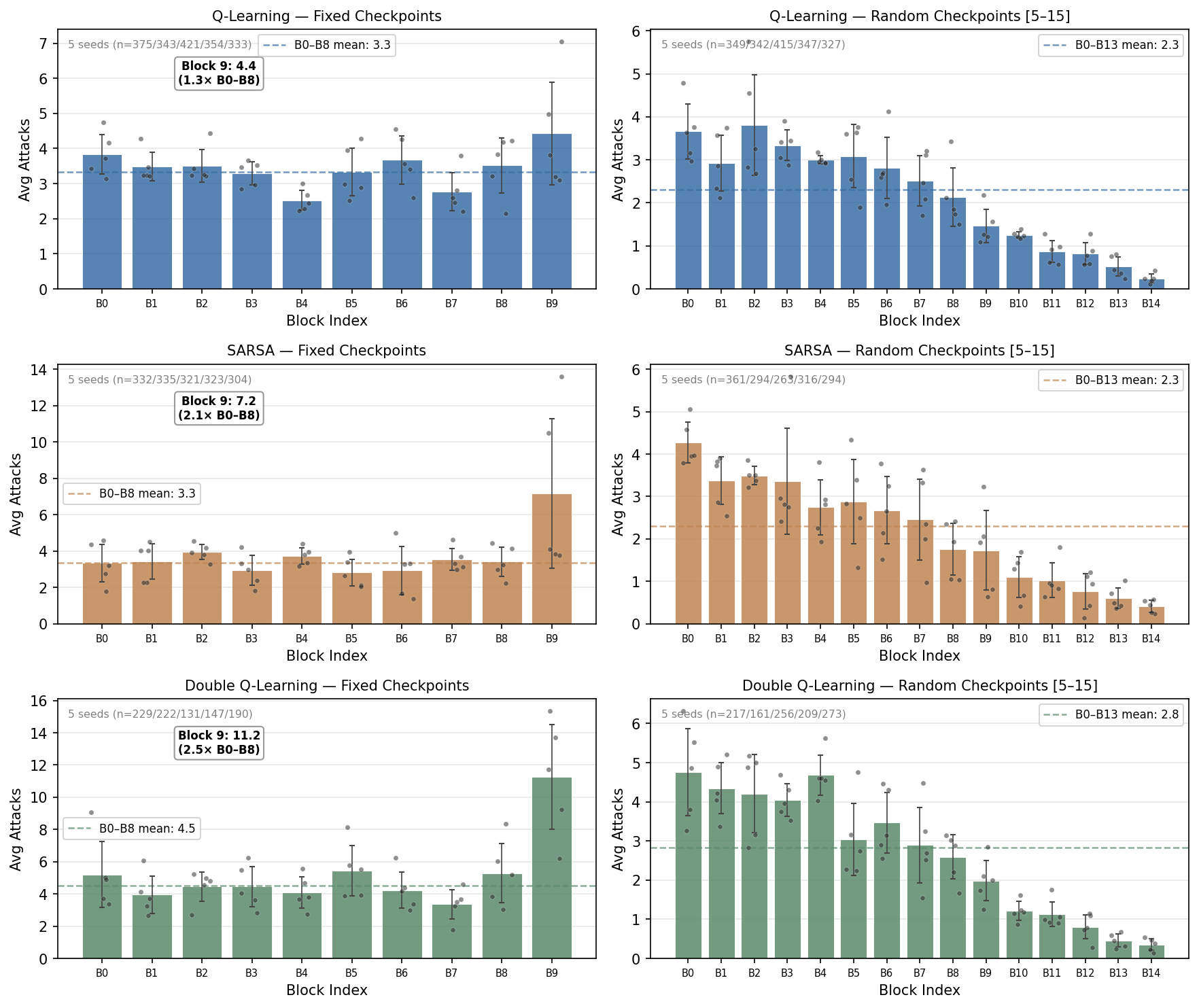}
\caption{Block-level attack profiles under fixed (left) and
random (right) checkpoints at $\mu_{ch}=3\%$. Bars show
5-seed means with $\pm$std error bars, and dark scatter points
overlay per-seed means. The terminal-block surge in the fixed
condition is replaced by a gradual decrease under randomized
horizons, which confirms the surge as a structural exploit of
the fixed-horizon design.}
\label{fig:burst_comparison}
\end{figure}

Under randomized checkpoints, the burst vanishes entirely
in all three agents (Fig.~\ref{fig:burst_comparison}).
Agents instead adopt the opposite pattern, where they attack more
aggressively early when accumulated QBER risk is low and
taper off as risk builds. Now we also ask whether
stealth performance changes under this randomization. Table~\ref{tab:extension} shows that detection rates under
random and fixed checkpoints are statistically
indistinguishable across all agents and noise levels. Eve
adapts her temporal strategy completely without any loss of
stealth effectiveness. This indicates that checkpoint
randomization eliminates the structural burst exploit while
leaving the underlying stealth capability intact.

\begin{table}[t]
\centering
\caption{Detection rate (\%) mean $\pm$ std: fixed vs.\
random checkpoint intervals, across 5 seeds.}
\label{tab:extension}
\small
\begin{tabular}{lcccccc}
\toprule
& \multicolumn{2}{c}{$\mu_{ch}=1\%$}
& \multicolumn{2}{c}{$\mu_{ch}=3\%$}
& \multicolumn{2}{c}{$\mu_{ch}=5\%$} \\
Agent & Fix. & Rnd. & Fix. & Rnd. & Fix. & Rnd. \\
\midrule
Q-Lrn. & $0.28$ & $0.36$ & $29.8$ & $28.8$ & $91.7$ & $90.6$ \\
SARSA  & $1.04$ & $3.68$ & $39.8$ & $38.9$ & $93.6$ & $89.3$ \\
DQ-Lrn.& $3.52$ & $2.00$ & $67.3$ & $55.4$ & $97.3$ & $96.8$ \\
\bottomrule
\end{tabular}
\end{table}

\section{Discussion}

We now discuss the implications of the results on what they reveal about the practical security margin
of BB84, why classical tabular RL is the appropriate tool
for this kind of analysis, and the limitations that bound
the scope of the conclusions. We also derive defensive
recommendations directly from the observed attack behaviors.

\subsection{What the Results Mean for BB84 Security}

Consider the security model assumed by most prior work where Eve
attacks at a fixed, pre-determined rate and Alice and Bob
monitor QBER against a fixed threshold. The results presented
here suggest that this model underestimates
the capability of a real adversary. An adaptive RL
attacker, one who observes the same QBER announcements as
the legitimate parties and adjusts accordingly, reduces
detection from the $\approx$100\% predicted by fixed-rate
analysis to $0.28\%\pm0.27\%$ at $\mu_{ch}=1\%$. The
security margin implied by the 11\% threshold is, in this
sense, much smaller in practice than in theory.

At $\mu_{ch}=1\%$, Q-Learning extracts 10.5 correct bits
per episode while remaining undetected. This leakage is
modest per session, but it accumulates across many sessions
without ever triggering the QBER monitoring system. Whether
this constitutes a practical key compromise depends on
deployment specifics, including session frequency and key reuse
policies.

The checkpoint randomization result adds an important
clarification. The end-game burst vanishes under
randomization, which is a genuine defensive improvement.
However, as shown in Table~\ref{tab:extension}, stealth does
not vanish with it. This observation points to a distinction
that matters for defense that the burst was an additional
exploit made possible by the fixed-horizon structure, and not
the foundation of the stealth capability. The root of the
vulnerability lies in the information asymmetry between an
adaptive, feedback-aware attacker and a detection mechanism
designed under a static threat model. Any countermeasure
that addresses only the burst without closing the stealth
window treats a symptom rather than the cause.

\subsection{Positioning of Classical RL on a Quantum Problem}

Let us be precise about what this work is and is not. We
do not propose a quantum algorithm, and we do not require
quantum hardware. The agents are standard classical tabular methods, applied to a classical simulation
of a quantum cryptographic protocol. The quantum physics
enters only through the simulation environment, specifically
through the relationship between attack rate and QBER. The
learning is entirely classical.

A consequence of this is that the demonstrated threat is
deployable today, without any advances in quantum computing.
Suppose an adversary has a physical tap on the quantum
channel, and access
to the publicly broadcast QBER announcements. The trained
policy can be developed offline in simulation and applied
at deployment. No quantum hardware is required at any stage.
This also means the results are fully reproducible without
specialized hardware, and the learned strategies are
interpretable in a way that quantum circuit-based policies
currently are not.

\subsection{Limitations}

Our agents operate against a non-adaptive Alice and Bob,
who use a fixed 11\% QBER threshold throughout. A more
realistic adversarial setting is one in which Alice and Bob
also adapt, for instance by dynamically adjusting the
threshold or varying block sizes. This could alter the stealth
window available to Eve in ways not captured here. Our
simulation also assumes perfect quantum optics. Photon
loss, dark counts, and detector inefficiency introduce noise
sources in real deployments that would affect the safe attack
budget and they are not included in our study.

These constraints are consistent with the proof-of-concept
scope of this work. Relaxing them by modeling adaptive
defenders, incorporating hardware noise, and scaling to
continuous state space represents a natural direction for
follow-on work. On the statistical side, the small sample
size of $N=5$ seeds limits power for detecting moderate
effect sizes. The comparison SARSA vs.\ Double Q-Learning
at $\mu_{ch}=3\%$ under random checkpoints ($p=0.094$) is
directionally consistent with the fixed-checkpoint result
but does not reach significance at the current sample size.
We report this transparently and replication with additional
seeds is a natural next step.

\subsection{Defensive Recommendations}

Three defensive measures follow from the experimental
findings. First is to randomize checkpoint intervals. This
eliminates the end-game burst and removes Eve's ability to
identify the terminal block. It does not close the stealth
window, but it removes one exploit at low implementation
cost. Second is to tighten QBER thresholds at low noise such as $\mu_{ch}=1\%$. The 11\% threshold leaves a stealth window
wide enough for a trained agent to operate indefinitely.
Reducing the threshold to $11\% - \delta$ for some carefully
chosen $\delta > 0$ would narrow this window, though this
must be balanced against increased false positive rates. Last is to model Eve as adaptive in security evaluations. The
gap between a fixed-rate Eve and an RL Eve is too large to
treat as a theoretical concern. Any rigorous BB84 security
assessment should include an adaptive adversary baseline
alongside the standard fixed-rate model.

\section{Conclusion}

We have shown that a classical tabular RL agent can learn to
eavesdrop on BB84 with near-perfect stealth at low channel
noise, where detection is reduced from 99.4\% to $0.28\%\pm0.27\%$
across five independent random seeds. This behavior is
statistically significant ($p=0.020$) and consistent across
three different algorithms whose behavioral differences map
onto their theoretical overestimation properties.

The end-game burst is the most structurally informative
finding. Agents learn, without being told, that the final
block of a fixed-horizon episode can be attacked freely
because detection there carries no future cost. All three
algorithms independently discover this pattern. When the
horizon is randomized, the burst vanishes. When we ask
whether stealth persists, it does, and at comparable rates.
The two phenomena are separable, and this separation
matters since randomizing checkpoints is a necessary but not
sufficient defense.

The broader implication follows from this separation. BB84
security evaluations that model Eve as a fixed-rate attacker
miss an entire class of adaptive exploits that require no
quantum computing to realize, only the QBER feedback that
the protocol already broadcasts. Treating the eavesdropper
as a learner, rather than a fixed policy, reveals a class of
vulnerabilities that standard security analyses are not
designed to detect.

We frame this work as a proof of concept rather than a
deployable attack. The agents are deliberately simple,
the simulation assumes ideal optics, and the defender is
non-adaptive. What the results establish is that adaptive
eavesdropping strategies are learnable in principle under
the BB84 reward structure using methods that predate modern
deep learning by three decades. Whether richer learning
methods, more sophisticated threat models, or hardware-based
simulations would amplify or attenuate the effect is left
to future work. The present contribution is to show that
the effect exists and is large enough to warrant attention.

\end{document}